\documentclass{aa}  

\usepackage{graphicx}
\usepackage{txfonts}
\usepackage{natbib}
\titlerunning{Transport of angular momentum in subgiant stars}

\begin{document}

   \title{Asteroseismology of evolved stars to constrain the internal transport of angular momentum}

   \subtitle{I. Efficiency of transport during the subgiant phase}

\author{P. Eggenberger\inst{1}
\and S. Deheuvels\inst{2}
\and A. Miglio\inst{3}
\and S. Ekstr\"om\inst{1}
\and C. Georgy\inst{1}
\and G. Meynet\inst{1}
\and N. Lagarde\inst{4}
\and S. Salmon\inst{5}
\and G.~Buldgen\inst{3,5}
\and J. Montalb\'{a}n\inst{6}
\and F. Spada\inst{7} 
\and J. Ballot\inst{2}}
\institute{Observatoire de Gen\`eve, Universit\'e de Gen\`eve, 51 Ch. des Maillettes, CH-1290 Sauverny, Suisse \\	
\email{patrick.eggenberger@unige.ch}
\and
IRAP, Universit\'e de Toulouse, CNRS, CNES, UPS, Toulouse, France
\and
School of Physics and Astronomy, University of Birmingham, Edgbaston, Birmingham B15 2TT, UK
\and
Institut Utinam, CNRS UMR 6213, Universit\'e de Franche-Comt\'e, OSU THETA Franche-Comt\'e-Bourgogne, Observatoire de Besan\c{c}on, 25010, Besan\c{c}on, France
\and
Département d'Astrophysique, Géophysique et Océanographie, Université de Liège, Allée du 6 Août 17, 4000 Liège, Belgium
\and
Dipartimento di Fisica e Astronomia, Universit\`a di Padova, Vicolo dell'Osservatorio 3, I-35122 Padova, Italy
\and
Max-Planck-Institut f\"ur Sonnensystemforschung, Justus-von-Liebig-Weg 3, 37077 G\"ottingen, Germany}
   \date{Received; accepted}

% \abstract{}{}{}{}{} 
% 5 {} token are mandatory
 
  \abstract
   {The observations of solar-like oscillations in evolved stars have brought important constraints on their internal rotation rates. To correctly reproduce these data, an efficient transport mechanism is needed in addition to the transport of angular momentum by meridional circulation and shear instability. The efficiency of this undetermined process is found to increase both with the mass and the evolutionary stage during the red giant phase.}
   {We study the efficiency of the transport of angular momentum during the subgiant phase.}
   {The efficiency of the unknown transport mechanism is determined during the subgiant phase by comparing rotating models computed with an additional corresponding viscosity to the asteroseismic measurements of both core and surface rotation rates for six subgiants observed by the \textit{Kepler} spacecraft. We then investigate the change in the efficiency of this transport of angular momentum with stellar mass and evolution during the subgiant phase.}
   {The precise asteroseismic measurements of both core and surface rotation rates available for the six \textit{Kepler} targets enable a precise determination of the efficiency of the transport of angular momentum needed for each of these subgiants. These results are found to be insensitive to all the uncertainties related to the modelling of rotational effects before the post-main sequence phase. An interesting exception in this context is the case of young subgiants (typical values of $\log(g)$ close to 4), because their rotational properties are sensitive to the degree of radial differential rotation on the main sequence. These young subgiants constitute therefore perfect targets to constrain the transport of angular momentum on the main sequence from asteroseismic observations of evolved stars. As for red giants, we find that the efficiency of the additional transport process increases with the mass of the star during the subgiant phase. However, the efficiency of this undetermined mechanism decreases with evolution during the subgiant phase, contrary to what is found for red giants. Consequently, a transport process with an efficiency that increases with the degree of radial differential rotation cannot account for the core rotation rates of subgiants, while it correctly reproduces the rotation rates of red giant stars. This suggests that the physical nature of the additional mechanism needed for the internal transport of angular momentum may be different in subgiant and red giant stars.}
   {}


   \maketitle
%
%-------------------------------------------------------------------

\section{Introduction}
\label{intro}

While the different dynamical processes related to rotation can potentially have a significant impact on the internal structure of a star, the exact modelling of such processes remains a major question of stellar evolution \citep[e.g.][]{mey13}. The possibility to reveal the internal rotation of stars represents a unique opportunity to better understand these transport mechanisms. This is now feasible thanks to asteroseismic observations of rotational frequency splittings for different kinds of stars.

In particular, observations by the \textit{Kepler} spacecraft \citep{bor10} of rotational splittings of mixed oscillation modes for subgiants and red giants have brought key information about the internal rotation of evolved stars \cite[][]{bec12, deh12, mos12, deh14, deh15, dim16, deh17, geh18}. Mixed modes behave similarly to pressure modes in the external stellar layers and to gravity modes in the center. They are thus especially valuable to simultaneously constrain the core and surface rotation rates of these stars, and hence the efficiency of the transport of angular momentum in stellar interiors. The comparison with rotating models of red giants has shown that hydrodynamic processes related to rotation, namely the transport of angular momentum by meridional circulation and shear instabilities, predict a high degree of radial differential rotation that is incompatible with asteroseismic rotation rates of red giants \citep{egg12_rg,cei13,mar13,can14}. Similarly to helioseismic measurements of the solar rotation profile, asteroseismic measurements of red giants point towards the need for an additional efficient mechanism for the transport of angular momentum in stellar radiative zones.

The physical nature of this additional transport process remains however unknown. Magnetic fields could ensure an efficient transport of angular momentum in stellar interiors. The inclusion of the Tayler-Spruit dynamo \citep{spr99,spr02} in rotating models leads indeed to an internal rotation in the solar radiative zone in good agreement with helioseismic measurements \citep{egg05_mag}. In the case of red giants, this dynamo is however found to provide an insufficient transport of angular momentum to correctly reproduce the low core rotation rates deduced from asteroseismic measurements \citep{can14}. In the same way, the transport of angular momentum by internal gravity waves is found to be inefficient in the central parts of red giants \citep{ful14,pin17} and is thus not able to account for the observed core rotation rates. By taking into account the generation of internal gravity waves by penetrative convection, \cite{pin17} have however shown that these waves could provide an efficient transport of angular momentum during the subgiant phase and may be able to correctly reproduce the internal rotation deduced from asteroseismic observations of subgiants. Conversely, the transport of angular momentum by mixed oscillation modes seems to be inefficient during the subgiant and the early red giant phase, while it could play an important role for the internal rotation of more evolved red giants \citep{bel15b}.

To improve our understanding of the physical nature of this missing transport process, one has to deduce its characteristics from asteroseismic measurements. The first step was to show that quantitative constraints on the efficiency of this unknown process can be directly obtained from asteroseismic data of red giants \citep{egg12_rg}. We then investigated how these constraints on the transport efficiency during the post-main sequence phase are sensitive to the rotational history of the star \citep{egg17}. We found that the efficiency of the internal transport of angular momentum during the post-main sequence can be precisely deduced from asteroseismic data of red giants independently from all uncertainties on the modelling of rotational effects on the main sequence (in particular regarding the modelling of the internal transport of angular momentum and surface magnetic braking during the main sequence). This is an important starting point to study how the efficiency of the unknown additional post-main sequence transport mechanism varies with the global stellar parameters. 

It has been found that the efficiency of the unknown process must increase when the star ascends the red giant branch \cite[][]{can14,egg15,spa16}. Interestingly, \cite{spa16} found that, for a fixed mass, a transport efficiency that increases with the radial rotational shear can correctly account for the core rotation rates obtained for red giants by \cite{mos12}. The efficiency of the internal transport of angular momentum also increases with the stellar mass for red giants at a similar evolutionary stage \citep{egg17}.

These characteristics of the unknown angular momentum transport mechanism have been deduced for stars on the red giant branch. In this paper, we focus on subgiant stars. The physics of the stellar models is described in Sect.~\ref{physics}. In Sect.~\ref{nuadd}, we study how the efficiency of the internal transport of angular momentum during the subgiant phase can be deduced from asteroseismic measurements of core and surface rotation rates. This is done by determining the additional viscosity needed for the six subgiant stars observed by \cite{deh14}. Section \ref{rothist} discusses the sensitivity of these results on the assumptions used for the rotational history of subgiant stars. The evolution of the transport efficiency during the subgiant phase is then discussed in Sect.~\ref{evonuadd} together with the consequences of these findings on the physical nature of the missing transport process. The conclusion is given in Sect.~\ref{conclusion}.

\section{Physics of the rotating models}
\label{physics}

This study builds upon, and is a continuation of the work devoted to the red giants KIC~8366239 and KIC~7341231 \citep{egg12_rg,egg17}. Models of subgiant stars are thus computed with the Geneva stellar evolution code \citep[{\scshape Genec},][]{egg08} by including rotational effects in the context of the assumption of shellular rotation \citep{zah92,mae98}. The evolution of the rotation profile is computed simultaneously to the evolution of the star by taking into account the internal transport of angular momentum by meridional currents and the shear instability \citep[see][for more details]{egg10_sl}. Similarly to the work done for red giants, we quantify the efficiency of the undetermined angular momentum transport mechanism by introducing an additional constant viscosity $\nu_{\rm add}$ corresponding to this process. This leads to the following equation for the internal transport of angular momentum in radiative zones:

\begin{equation}
  \rho \frac{{\rm d}}{{\rm d}t} \left( r^{2}\Omega \right)_{M_r} 
  =  \frac{1}{5r^{2}}\frac{\partial }{\partial r} \left(\rho r^{4}\Omega
  U(r)\right)
  + \frac{1}{r^{2}}\frac{\partial }{\partial r}\left(\rho (D_{\rm shear}+\nu_{\rm add}) r^{4}
  \frac{\partial \Omega}{\partial r} \right) \, . 
\label{transmom}
\end{equation}

\noindent $r$ and $\rho$ correspond to the characteristic radius and mean density on an isobar. $\Omega(r)$ is the mean angular velocity and $U(r)$ denotes the velocity of meridional currents in the radial direction. $D_{\rm shear}$ is the diffusion coefficient associated to the transport of angular momentum by the shear instability and $\nu_{\rm add}$ is the additional viscosity corresponding to the unknown transport process. A very efficient transport of angular momentum leading to a flat rotation profile is assumed in convective zones. This is motivated by helioseismic measurements, which reveal a low degree of differential rotation in the radial direction in the solar convective envelope. 

We recall here that we obviously do not expect the undetermined additional mechanism of angular momentum transport to be effectively described by the simple assumption of a constant viscosity in stellar radiative zones. The key idea of introducing such an additional mean viscosity is first to be able to study whether asteroseismic measurements of evolved stars can really provide us with quantitative constraints on the internal transport of angular momentum during the post-main sequence phase without being sensitive to the uncertainties related to the modelling of rotational effects on the main sequence. We have shown that this is effectively the case for red giant stars \citep{egg17}, and we are now interested in determining whether this is also the case for less evolved stars like the subgiants. The second key point that can be addressed by introducing such an additional viscosity is the variation of the mean efficiency of the unknown transport process with global stellar parameters. Comparisons of asteroseismic data with rotating models of red giants show that the transport efficiency must increase both with the mass of the star and its evolution on the red giant branch \citep{egg17}. Are such trends also seen during the subgiant phase?
 
To characterize the internal transport of angular momentum during the subgiant phase, we compare rotating models to asteroseismic measurements obtained for six subgiant stars by \cite{deh14}. Contrary to the case of the more evolved red giant stars for which only an upper limit on the surface rotation rate is usually deduced from rotational splittings of mixed modes \citep[see however][for a new method to constrain the surface rotation of red giants]{deh17}, a precise determination of the surface rotation rate is available in addition to the core rotation rate for these six subgiants. This is of course particularly valuable to constrain the efficiency of the internal transport of angular momentum.

\begin{figure}[htb!]
\resizebox{\hsize}{!}{\includegraphics{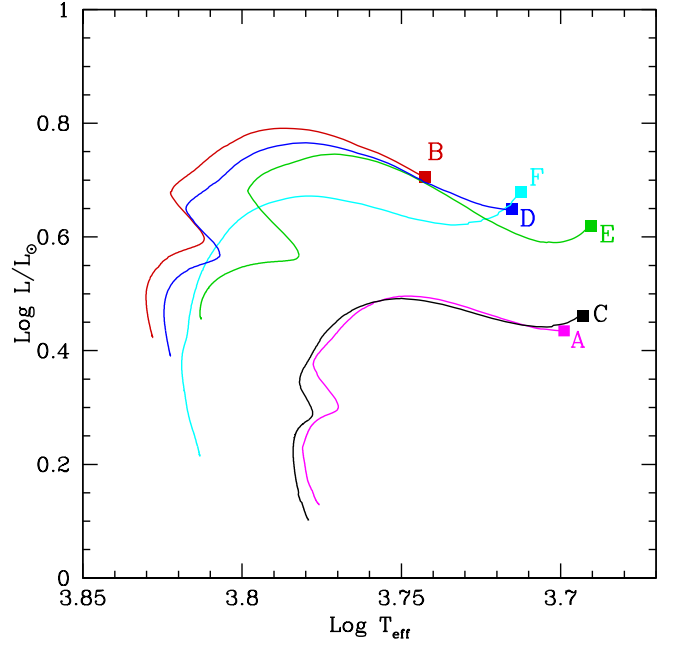}}
 \caption{Evolutionary tracks in the HR diagram for rotating models of the six subgiants observed by \cite{deh14}. The letters refer to the models shown in Table~\ref{tab1}.}
  \label{dhr_sub}
\end{figure}

An individual modelling is performed for each of these six subgiants by computing rotating models that takes into account the transport of angular momentum as described in Eq.~\ref{transmom}. Similarly to the work done for the low-mass red giant KIC~7341231 \citep{cei13,egg17}, a rotating model that simultaneously reproduces the observed values of the large frequency separation and the asymptotic period spacing is considered as a representative model of a subgiant. We use the solar chemical mixture of \cite{gre93} and the mixing-length parameter is fixed to its solar-calibrated value. To constrain the initial rotation velocity of the models, we use the surface rotation rates deduced from asteroseismic data for the six subgiants \citep[Table 13 of][]{deh14}. We then compute a small grid of rotating models with different initial velocities on the zero-age main sequence (ZAMS) for each of the six subgiants. As a first step, the transport of angular momentum in radiative zones is assumed to be only due to meridional currents and shear instability, and no braking of the surface by magnetized winds is included in the computation. The latter assumption, which is of course doubtful regarding the observations of surface rotation velocities for solar-type stars on the main sequence, can be first justified by recalling that the modelling of the surface braking has no impact on the determination of the efficiency of the internal transport of angular momentum for red giants \citep{egg17}. It remains however to be seen whether such a conclusion is also valid for subgiant stars; this is studied in Sect.~\ref{rothist}. At the end, we obtain representative rotating models of the six subgiants computed with the {\scshape Genec} code. The initial velocities on the ZAMS of these models are respectively equal to 6, 5, 5, 6, 8 and 4\,km\,s$^{-1}$ for stars A, B, C, D, E and F. These low initial velocities are a direct consequence of the assumption of an inefficient braking of the stellar surface during the main sequence. The evolutionary tracks of these models are shown in Fig.~\ref{dhr_sub}.
The input parameters and the global properties of these models are given in Table~\ref{tab1} together with the ones obtained by \cite{deh14} with the {\scshape Cesam2k} code. The characteristics of these rotating models, and in particular the masses, are very similar to the best-fit non-rotating models found by \cite{deh14}.

\begin{table*}
\caption{Input parameters and global properties of the rotating models computed in the present work with the {\scshape Genec} code. The properties of the {\scshape Cesam2k} models and the values of the ratio of core to surface rotation rates are taken from \cite{deh14}.}
\begin{center}
\begin{tabular}{l c | c c c | c c c | c c}
\hline \hline
 & Evolution code & $M/M_\odot$ & $(Z/X)_{\rm ini}$ & $Y_{\rm ini}$ & age [Gyr] & $T_{\rm eff}$ [K] & $\log g$ & $\Omega_{\rm c}/\Omega_{\rm s}$ & $\nu_{\rm add}$ [cm$^2$\,s$^{-1}$] \\
\hline
Star A  & {\scshape Genec} &	1.20 & 	0.0550 &  0.30 & 	6.7 &  	4986 & 	3.823 & 	$2.5 \pm 0.7$ & 	$(1.5 \pm 0.5) \times 10^{4}$ \\
    & 	{\scshape Cesam2k} &	1.22 & 	0.0500 &  0.30 &	5.9 &  	5026 & 	3.826 &  &  \\
% & & & & & & & & & \\
\hline
Star B  & {\scshape Genec} &	1.27 & 	0.0190 &  0.28 &	3.7 & 	5511 & 	3.765 & 	$3.8 \pm 1.1$ & 	$(1.7 \pm 0.6) \times 10^{4}$ \\
 & 	{\scshape Cesam2k} &	1.27 &  0.0173 &  0.27 & 	3.8 & 	5575 & 	3.758 &  & 	 \\
\hline
Star C  & {\scshape Genec} &	1.15 & 	0.0390 &  0.29 & 	7.3 &  	4966 & 	3.771 & 	$6.9 \pm 1.0$ & 	$(6.0 \pm 1.5) \times 10^{3}$ \\
    & 	{\scshape Cesam2k} &	1.14 & 	0.0388 &  0.30 &	6.9 &  	4973 & 	3.772 &  &  \\
\hline
Star D  & {\scshape Genec} &	1.25 & 	0.0160 &  0.26 &	4.3 & 	5186 & 	3.685 & 	$15.1 \pm 4.0$ & 	$(6.0 \pm 2.0) \times 10^{3}$ \\
 & 	{\scshape Cesam2k} &	1.26 &  0.0151 &  0.27 & 	3.8 & 	5281 & 	3.685 &  & 	 \\
\hline
Star E  & {\scshape Genec} &	1.40 & 	0.0500 &  0.29 & 	3.9 &  	4927 & 	3.673 & 	$10.7 \pm 1.5$ & 	$(1.0 \pm 0.2) \times 10^{4}$ \\
    & 	{\scshape Cesam2k} &	1.39 & 	0.0548 &  0.30 &	3.8 &  	4898 & 	3.677 &  &  \\
\hline
Star F  & {\scshape Genec} &	1.10 & 	0.0100 &  0.26 &	5.9 & 	5145 & 	3.579 & 	$21.0 \pm 8.2$ & 	$(4.0 \pm 2.0) \times 10^{3}$ \\
 & 	{\scshape Cesam2k} &	1.07 &  0.0116 &  0.27 & 	6.0 & 	5194 & 	3.574 &  & 	 \\
\hline
\end{tabular}
\end{center}
\label{tab1}
\end{table*}

\begin{figure}[htb!]
\resizebox{\hsize}{!}{\includegraphics{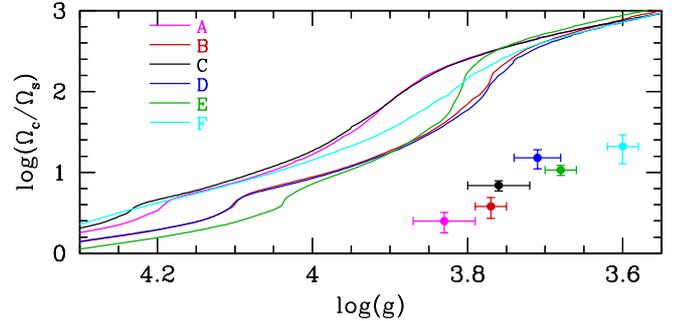}}
 \caption{Ratio of core to surface rotation rates as a function of gravity for rotating models of the six subgiants. These models are computed without additional viscosity ($\nu_{\rm add}=0$) and an initial velocity on the ZAMS of 6, 5, 5, 6, 8 and 4\,km\,s$^{-1}$ for stars A, B, C, D, E and F, respectively. Dots indicate the values of the ratio of core to surface rotation rates determined for the six subgiants from asteroseismic measurements \citep{deh14}.}
  \label{rap_omegasc_seb_noDadd}
\end{figure}

While these rotating models correctly reproduce the observed surface rotation rates of the six subgiant stars, they predict rapidly rotating cores in contradiction with asteroseismic observations. This is shown in Fig.~\ref{rap_omegasc_seb_noDadd} by comparing the ratio of core to surface rotation rates of these models to the observed values deduced by \cite{deh14}. Theoretical values are found to be more than one order of magnitude larger (typically between 20 and 80 times larger) than observed for these subgiant stars. As for red giants, we conclude that meridional currents and shear instabilities alone do not provide a sufficient coupling to correctly account for the core rotation rates observed during the subgiant phase. An efficient additional mechanism for the transport of angular momentum in stellar radiative zones is thus also required for these subgiant stars.

\section{Additional viscosity during the subgiant phase}
\label{nuadd}

\subsection{Determination of the additional viscosity}

To quantify the efficiency of the required additional process for the internal angular momentum transport, we compute rotating models for each of the six subgiants using different values of the additional viscosity $\nu_{\rm add}$ (see Eq.~\ref{transmom}). The values of $\nu_{\rm add}$ are then constrained by requiring that these models correctly reproduce the observed ratio of core to surface rotation rates. Contrary to the work done for the red giant KIC~7341231 for which only an upper limit on the surface rotation rate was available, a precise estimate of the surface rotation rate can be deduced from asteroseismic data for these less evolved subgiants. This enables a precise determination of the efficiency of the additional transport process needed during the subgiant phase.

\begin{figure}[htb!]
\resizebox{\hsize}{!}{\includegraphics{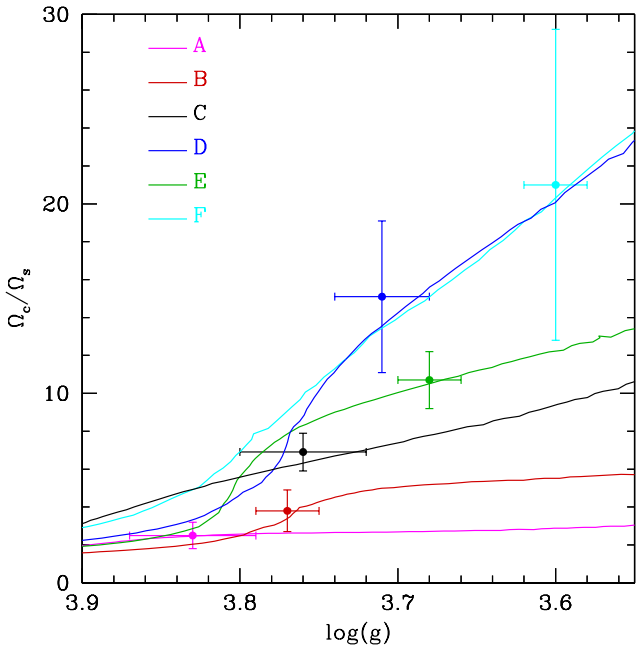}}
 \caption{Ratio of core to surface rotation rates as a function of gravity for rotating models of the six subgiants. These models are computed with an additional viscosity $\nu_{\rm add}$ of $1.5 \times 10^{4}$, $1.7 \times 10^{4}$, $6 \times 10^{3}$, $6 \times 10^{3}$, $1 \times 10^{4}$ and $4 \times 10^{3}$\,cm$^2$\,s$^{-1}$ for stars A, B, C, D, E and F, respectively. Dots indicate the values of the ratio of core to surface rotation rates determined for the six subgiants from asteroseismic measurements \citep{deh14}.}
  \label{rap_omegacs_seb}
\end{figure}

\begin{figure}[htb!]
\resizebox{\hsize}{!}{\includegraphics{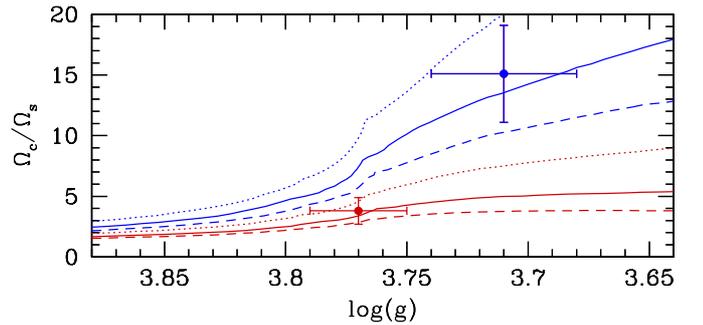}}
 \caption{Ratio of core to surface rotation rates as a function of gravity for rotating models of stars B (red) and D (blue). For star B, dotted, solid and dashed lines correspond to an additional viscosity $\nu_{\rm add}$ of $1.1 \times 10^{4}$, $1.7 \times 10^{4}$ and $2.3 \times 10^{4}$\,cm$^2$\,s$^{-1}$. For star D, dotted, solid and dashed lines correspond to an additional viscosity $\nu_{\rm add}$ of $4 \times 10^{3}$, $6 \times 10^{3}$ and $8 \times 10^{3}$\,cm$^2$\,s$^{-1}$.}
  \label{rap_omegacs_seb_BD}
\end{figure}

The values of the mean additional viscosity $\nu_{\rm add}$ obtained for the six subgiants are given in Table~\ref{tab1}. Figure~\ref{rap_omegacs_seb} shows the evolution of the ratio of core to surface rotation rates as a function of gravity for these models. The uncertainty on the value of $\nu_{\rm add}$ can be simply estimated from the uncertainty on the observed ratio of core to surface rotation rates. This is illustrated in Fig.~\ref{rap_omegacs_seb_BD} for two different subgiants: a star with a low value of the ratio of core to surface rotation rate (star B), and a more evolved subgiant characterized by a larger value of this ratio (star D). The degree of radial differential rotation decreases when $\nu_{\rm add}$ increases because of the more efficient transport of angular momentum. As a result, the range of possible values for the additional viscosity $\nu_{\rm add}$ can be constrained by requiring that the models correctly reproduce the observed ratio of core to surface rotation rates.

\begin{figure}[htb!]
\resizebox{\hsize}{!}{\includegraphics{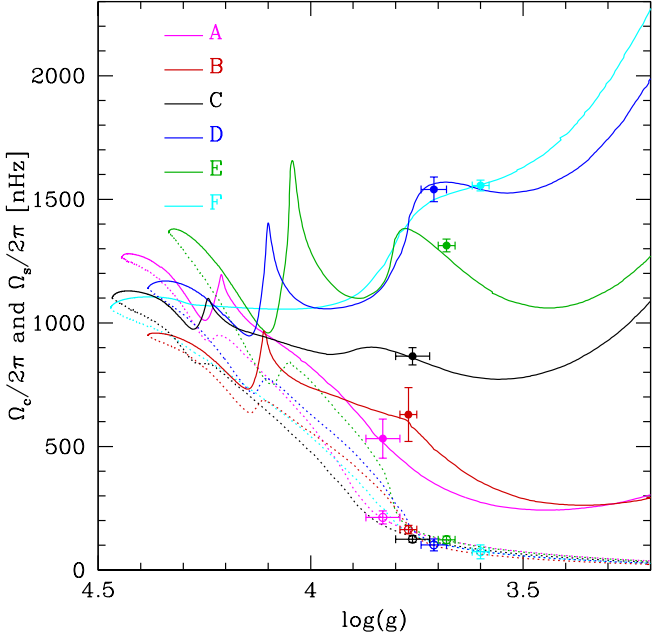}}
 \caption{Core and surface (solid and dotted lines) rotation rates as a function of gravity for rotating models of the six subgiants. These models are computed with an additional viscosity $\nu_{\rm add}$ of $1.5 \times 10^{4}$, $1.7 \times 10^{4}$, $6 \times 10^{3}$, $6 \times 10^{3}$, $1 \times 10^{4}$ and $4 \times 10^{3}$\,cm$^2$\,s$^{-1}$ for stars A, B, C, D, E and F, respectively. The initial velocity on the ZAMS is equal to 6, 5, 5, 6, 8 and 4\,km\,s$^{-1}$ for stars A, B, C, D, E and F, respectively. Dots and open circles indicate the values of core and surface rotation rates determined for the six subgiants from asteroseismic measurements \citep{deh14}.}
  \label{omegacs_vs_logg_seb}
\end{figure}

As found for the red giant KIC~7341231, the value of the ratio of the core to surface rotation rate is almost insensitive to the adopted initial velocity on the ZAMS. Of course, a given value of the additional viscosity constrains the initial velocity on the ZAMS in order to predict surface and core rotation rates compatible with the ones deduced from asteroseismic measurements. The evolution of the core (solid lines) and surface (dotted lines) rotation rates for the same models shown in Fig.~\ref{rap_omegacs_seb} is plotted in Fig.~\ref{omegacs_vs_logg_seb}. We recall that the low values found for the initial velocities on the ZAMS for these models (between 4 and 8\,km\,s$^{-1}$) are directly due to the neglect of the braking of the stellar surface by magnetized wind. The impact of this assumption on the determination of the efficiency of the internal transport of angular momentum in subgiants is studied in Sect.~\ref{rothist}.

The values obtained for the additional viscosity $\nu_{\rm add}$ of the six subgiants (see Table~\ref{tab1}) are of the same order as the ones already determined for red giants. In particular, these values lie in-between the mean efficiency of $3 \times 10^{4}$\,cm$^2$\,s$^{-1}$ found for the 1.5\,$M_{\odot}$ red giant KIC~8366239 and the lower value of about $(1-4)\times 10^{3}$\,cm$^2$\,s$^{-1}$ found for the 0.84\,$M_{\odot}$ red giant KIC~7341231. Recalling that the six subgiants are characterized by masses in-between the ones of these two red giants, this result is compatible with the increase of the efficiency of the internal transport of angular momentum with the stellar mass. However, this simple comparison does not take into account differences in other global stellar properties, and in particular the metallicity. This is important in the case of the red giant KIC~7341231, which is characterized by a lower metallicity than the six subgiants studied here ($[$Fe/H$] = -1$ for KIC~7341231). As investigated in Sect.~5 of \cite{egg17}, the metallicity seems also to have an impact on the efficiency needed for the additional transport mechanism: this preliminary study suggests that a decrease in the metallicity would result in a higher efficiency needed for the unknown transport mechanism. The mean efficiency of $(1-4)\times 10^{3}$\,cm$^2$\,s$^{-1}$ determined for the 0.84\,$M_{\odot}$ low-metallicity red giant KIC~7341231 is then expected to be higher than the efficiency that would be obtained for a star with the same global parameters (in particular the same mass) but a solar chemical composition. The increase of the efficiency of the transport of angular momentum with the stellar mass seems thus to be visible here, despite the different metallicities of the stars considered.

\subsection{Evolutionary sequence of the six subgiants}

Comparing the values determined for $\nu_{\rm add}$ with the gravity of the six subgiants, one does not see a clear trend. The highest values of $\nu_{\rm add}$ seem simply to be preferentially found for subgiants with higher values of gravity, but without a perfect match with the surface gravity (we recall that the six subgiants are named according to their values of surface gravity, with star A characterized by the highest value of $\log(g)$). According to the efficiency of the internal transport of angular momentum, the following sequence is indeed found (for a decreasing value of $\nu_{\rm add}$): B, A, E, C/D, F. Based on the evolutionary tracks shown in Fig.~\ref{dhr_sub}, the following evolutionary order is suggested (from the less to the most evolved subgiant): B, A, D, C, E, F. Interestingly, this evolutionary sequence seems to be similar to the sequence B, A, E, C/D, F deduced from the efficiency of the transport of angular momentum (except of course of star E). This suggests that the efficiency of the additional transport process must decrease when the star evolves during the subgiant phase. 

\begin{figure}[htb!]
\resizebox{\hsize}{!}{\includegraphics{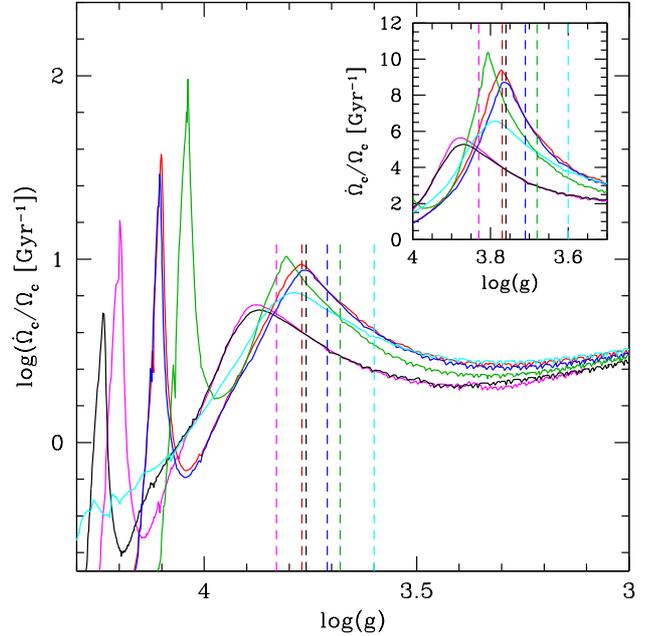}}
 \caption{Variation rate of the core angular velocity $\Omega_{\rm c}$ as a function of gravity for rotating models of the six subgiants. These models are computed without an additional transport of angular momentum ($\nu_{\rm add}=0$). Magenta, red, black, blue, green and cyan lines correspond to stars A, B, C, D, E and F, respectively. Vertical dashed lines indicate the values of the gravity for the six stars. A zoom during the subgiant phase is shown in the insert.}
  \label{domegac_omc_dt_seb}
\end{figure}

To study this in more detail, we introduce the variation rate of the core angular velocity $\dot{\Omega}_{\rm c}/\Omega_{\rm c}$, which is the inverse of the characteristic timescale of central contraction. This variation rate constitutes a valuable indicator of the evolutionary stage of a subgiant star from the point of view of its rotational properties. The evolution of the variation rate of the core angular velocity as a function of gravity is shown in Fig.~\ref{domegac_omc_dt_seb} for the six subgiants. The tracks in this figure correspond to models with $\nu_{\rm add}=0$. This enables to quantify the capability of a star to create radial differential rotation despite the transport of angular momentum by meridional currents and shear instability, and hence to have a first idea of the efficiency needed by an additional transport mechanism to counteract the related increase of the core rotation rate. Note that the transport of angular momentum by meridional currents and shear instability is almost negligible during the subgiant phase due to the rapid evolution that characterized this phase \cite[see e.g. Fig. 7 of][]{cei13}. 

Figure~\ref{domegac_omc_dt_seb} illustrates the typical behavior of the variation rate of the core angular velocity during the end of the main sequence and the post-main sequence phase. The evolution of this quantity is first characterized by a sharp peak during the end of the evolution on the main sequence (the end of the main sequence is defined here by a central mass fraction of hydrogen $X_{\rm c}$ lower than $10^{-5}$) and a second broader peak in the middle of the subgiant phase. The first peak is due to the overall contraction at the end of the main sequence. This global contraction leads to a decrease of the stellar radius and gives rise to the characteristic hook-like feature of the evolutionary tracks at the end of the main sequence with both an increase of the effective temperature and luminosity of the star (see Fig.~\ref{dhr_sub}). This results in the simultaneous increase of the core and surface rotation rates, as shown in Fig.~\ref{omegacs_vs_logg_seb}. This rapid contraction is only at work for stars with a convective core during the main sequence. This is the case for all subgiants of our sample except star F. The low mass and metallicity of this latter star lead indeed to a radiative core during the main sequence and hence to a slow contraction of the central layers at the end of the main sequence. No hook-like feature is then seen for this star in the HR diagram (see Fig.~\ref{dhr_sub}) and no sharp peak of the variation rate of the core angular velocity is present at the end of the main sequence (see Fig.~\ref{domegac_omc_dt_seb}).

The second broader peak of the variation rate of the core angular velocity seen in Fig.~\ref{domegac_omc_dt_seb} characterizes the contraction of the central layers during the subgiant phase. The amplitude of this peak is directly related to the capability of the star to create radial differential rotation (a higher amplitude meaning a more rapid increase of the core rotation rate). Moreover, the location of the star relative to this peak offers a possibility to estimate its evolutionary stage during the subgiant phase from a rotational point of view. As seen in the inset of Fig.~\ref{domegac_omc_dt_seb}, star B is located at the maximum of the peak, while all other stars are found after this maximum. Star B is thus the least evolved star and the following evolutionary sequence can then be deduced from the variation rate of the core angular velocity (from the less to the most evolved subgiant): B, A, D, C, E, F. This evolutionary sequence is the same as the one estimated above from the evolutionary tracks in the HR diagram shown in Fig.~\ref{dhr_sub}. This confirms that, once the specific masses and metallicities of the subgiants are taken into account, the evolutionary sequence of these targets is B, A, D, C, E, F instead of the sequence A to F, which is simply defined by the decrease of the surface gravity.

\subsection{Evolution of the transport efficiency}

The sequence deduced from the decreasing efficiency of internal angular momentum transport (i.e. B, A, E, C/D, F) follows the evolutionary sequence B, A, D, C, E, F (exception made of star E). This decrease of the transport efficiency during the subgiant evolution can also be illustrated by comparing stars with approximately the same mass and chemical composition but different evolutionary stages. This is the case of stars B and D: a value of the additional viscosity three times lower is then obtained for the more evolved star D. In the same way, models of stars A and C share similar initial input parameters, and a lower efficiency of the missing transport process is determined for the more evolved star C. 

As for red giants, an increase of the efficiency of the additional transport mechanism with the stellar mass is found during the subgiant phase. This can be seen by comparing stars located at a similar evolutionary stage that have different masses. This is the case of stars C and E: as shown in Table~\ref{tab1} an increase of the mass leads to an increase of the needed mean additional viscosity. A word of caution is however needed here, because these stars also exhibit different metallicities. As mentioned above, a change in metallicity can impact the efficiency determined for the additional transport process. Preliminary results obtained for red giants indicate that an increase in metallicity leads to a decrease in the additional viscosity needed \citep{egg17}. For the specific case of stars C and E, the metallicity is higher for the more massive star. Thus, the increase of the additional viscosity with the stellar mass seen for these subgiants cannot be attributed to differences in metallicites. In particular, the dependence of the additional viscosity on the stellar mass explains the high value of $\nu_{\rm add}$ found for star E. This star is indeed characterized by a significantly higher mass ($1.4$\,$M_{\odot}$) compared to the other subgiants (with masses closer to $1.2$\,$M_{\odot}$). Consequently, star E has also a significantly higher value of the additional viscosity than other subgiants sharing a similar evolutionary stage. This explains why star E does not follow the simple sequence of decreasing efficiency of angular momentum transport as stars evolve during the subgiant phase.

\begin{figure}[htb!]
\resizebox{\hsize}{!}{\includegraphics{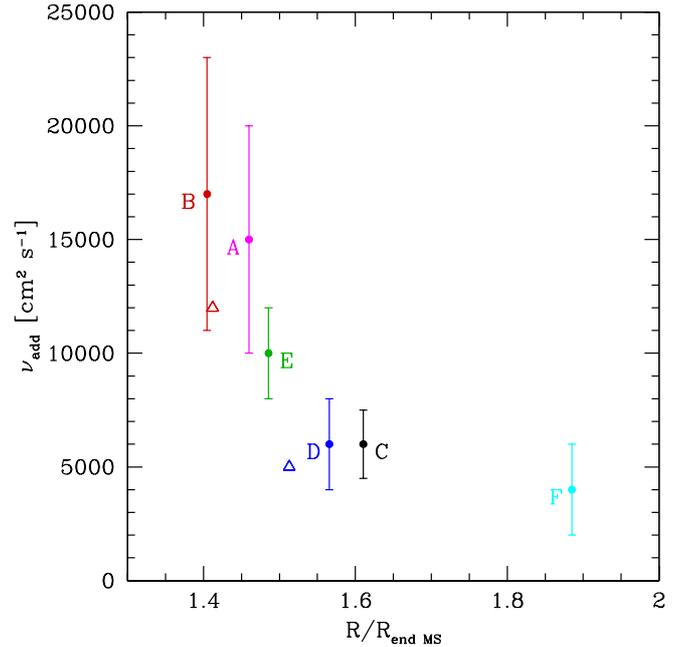}}
 \caption{Viscosity corresponding to the additional mechanism for the internal transport of angular momentum during the subgiant phase as a function of the radius of the star normalized to the radius at the end of the main sequence. Dots indicate the values determined in the present study. Open triangles correspond to the values of the 1.25\,$M_{\odot}$ models studied in \cite{spa16}, which are used as approximate models for stars B and D (red and blue triangles respectively, see text for more details).}
  \label{nuadd_vs_RRms}
\end{figure}

\begin{figure}[htb!]
\resizebox{\hsize}{!}{\includegraphics{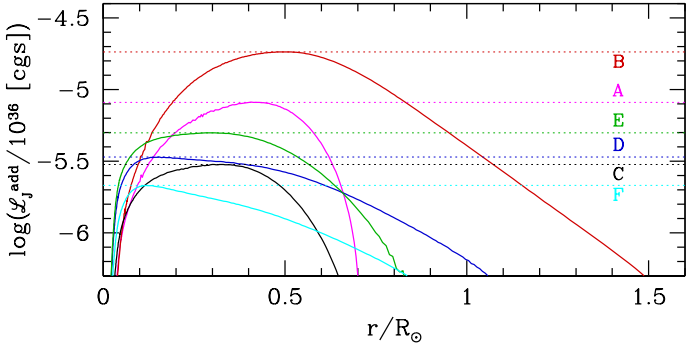}}
 \caption{Angular momentum luminosity as a function of the radius in solar units (${\cal L}_{J}^{\rm add}=-4\pi\rho \nu_{\rm add} r^{4} \frac{\partial \Omega}{\partial r}$) associated to the additional mechanism needed for the transport of angular momentum in the radiative zones of subgiant stars. Horizontal dotted lines indicate the maximum value of this luminosity for the six subgiants.}
  \label{lumin_flux_fig}
\end{figure}

The evolution of the efficiency of the unknown transport mechanism during the subgiant phase is illustrated in Fig.~\ref{nuadd_vs_RRms}. This figure shows the variation of the additional viscosity determined for each of the six subgiants as a function of the radius of the star normalized to the radius of the corresponding model at the end of the main sequence. The decrease of the efficiency of the internal transport of angular momentum with the evolution of the star during the subgiant phase is then clearly visible. In addition to the values obtained in the present study, Fig.~\ref{nuadd_vs_RRms} also shows the viscosity estimated for stars B and D (red and blue open triangles in Fig.~\ref{nuadd_vs_RRms}) from models computed with the YREC evolution code by \cite{spa16}. These models were computed with a fixed mass of 1.25\,$M_{\odot}$ and a solar chemical composition \citep[see Fig.~2 of ][]{spa16}. They do not correspond to exact models of stars B and D, but their input parameters are not too different from the ones determined for these subgiants. These estimates are compatible with the values determined in the present study. This indicates that the result of a decreasing efficiency of the transport with the evolution during the subgiant phase is not too sensitive to the exact values of the global parameters of the best-fit models determined for the subgiants. This also suggests that the present results are not too sensitive to the stellar evolution code used. 

We finally show in Fig.~\ref{lumin_flux_fig} that the decrease in the transport efficiency is not limited to the value of the needed additional viscosity, but corresponds to a global decrease of the angular momentum luminosity ${\cal L}_{J}^{\rm add}$ associated to the additional transport process. We thus conclude that, contrary to the case of red giants, the efficiency of the unknown additional mechanism for the internal transport of angular momentum decreases with the evolution of the star during the subgiant phase.

\section{Effects of the rotational history of subgiant stars on the determination of the additional viscosity}
\label{rothist}

The efficiency of the additional transport mechanism needed during the post-main sequence phase has been determined in the preceding section by assuming that the same process is already at work during the main sequence and by neglecting the possible impact of magnetized winds on the braking of the stellar surface. We now address the key question of the sensitivity of the determination of the additional viscosity needed during the subgiant phase on the rotational history of the star. 

To investigate this point, we compute models of subgiants with different assumptions regarding the transport of angular momentum in the radiative zones during the main sequence. We also compute models with a braking of the stellar surface by magnetized winds. These models are computed with the braking law of \cite{kri97} with a critical saturation value $\Omega_{\rm sat}$ fixed to 8\,$\Omega_{\odot}$ and a braking constant $K$ fixed to 1, 1/2 and 1/3 of its solar-calibrated value \citep[see Sect.~3 of][for more details]{egg17}. Solar and subsolar values of the braking constant $K$ are used for these subgiants, since a decrease of $K$ is expected when the mass of the star increases \citep[see e.g. Table 3 of][]{lan15}.

\begin{figure}[htb!]
\resizebox{\hsize}{!}{\includegraphics{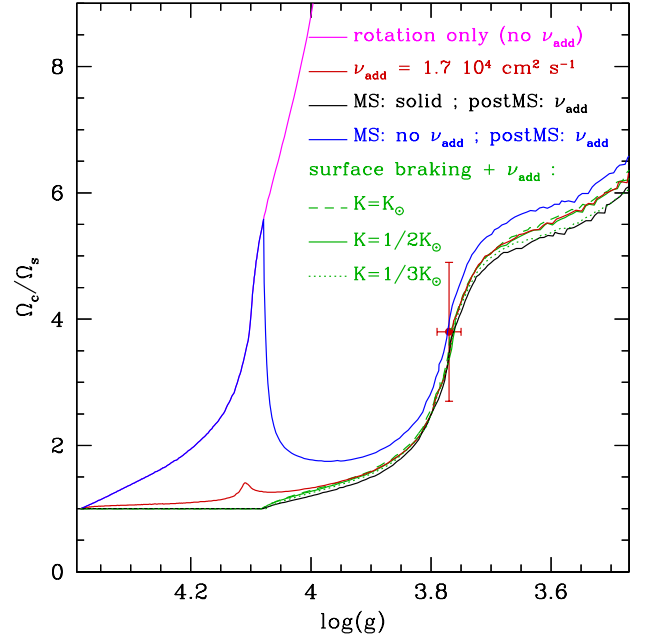}}
 \caption{Ratio of core to surface rotation rates as a function of gravity for rotating models of star B. The magenta line corresponds to a model without additional viscosity. Black and blue lines indicate models computed with solid-body rotation and without additional viscosity during the main sequence, respectively. Both models include an additional viscosity $\nu_{\rm add}$ of $1.7 \times 10^{4}$\,cm$^2$\,s$^{-1}$ during the post-main sequence. The red line indicates a model with an additional viscosity $\nu_{\rm add}=1.7 \times 10^{4}$\,cm$^2$\,s$^{-1}$ during the whole evolution. All these models are computed without braking of the surface by magnetized winds. The green lines correspond to the same models as the black line except for the inclusion of surface magnetic braking: dashed, continuous and dotted lines correspond to a braking constant $K$ equal to 1, 1/2 and 1/3 of its solar-calibrated value, respectively. The red dot indicates the ratio of core to surface rotation rates determined for star B by \cite{deh14}.}
  \label{rap_omegacs_seb_B}
\end{figure}

The effects of these various assumptions on the ratio of core to surface rotation rates of the models are illustrated in Fig.~\ref{rap_omegacs_seb_B} for star B. The example of this star is particularly interesting to consider, because it is the less evolved subgiant of the sample (see discussion above) and is thereby the most susceptive to be influenced by the modelling of rotation during the main sequence. The first model shown in Fig.~\ref{rap_omegacs_seb_B} is computed without additional viscosity during the whole evolution; it only includes the internal transport of angular momentum by meridional currents and shear instability (magenta line). Such a model predicts a very high degree of differential rotation after the main sequence as a result of the rapid increase of the core rotation rate (predicted value approximately 40 times higher than the core rotation rate observed for star B). When an additional viscosity of $1.7 \times 10^{4}$\,cm$^2$\,s$^{-1}$ is included in the computation, the asteroseismic value of the core to surface rotation rates of star B is correctly reproduced: this is the model of star B discussed in the preceding section (red line in Fig.~\ref{rap_omegacs_seb_B}). 

The black line in Fig.~\ref{rap_omegacs_seb_B} shows the effects of a very efficient transport of angular momentum during the main sequence on the value of the additional viscosity needed during the subgiant phase. This is done by using the extreme case of solid-body rotation for the evolution before the post-main sequence phase (i.e. when $X_{\rm c}$ is higher than $10^{-5}$) and then the additional viscosity required during the subgiant phase is determined. The black line in Fig.~\ref{rap_omegacs_seb_B} corresponds to an additional viscosity of $1.7 \times 10^{4}$\,cm$^2$\,s$^{-1}$ after the main sequence. A rapid convergence towards the rotation rates of the model computed without the assumption of solid-body rotation on the main sequence (red line) is seen during the subgiant phase.

Reducing the efficiency of the internal transport of angular momentum during the main sequence leads to higher values of the core rotation rate and lower values of the surface rotation rate at the beginning of the post-main sequence evolution. This is shown by the blue line in Fig.~\ref{rap_omegacs_seb_B}, which correspond to a model computed without additional viscosity during the main sequence and a value of $\nu_{\rm add}=1.7 \times 10^{4}$\,cm$^2$\,s$^{-1}$ after the main sequence. This increase of the radial differential rotation is seen during the whole evolution on the main sequence, but becomes particularly significant close to its end, due to the rapid overall contraction of the star (compare the blue and red lines in Fig.~\ref{rap_omegacs_seb_B}). One would then predict that this higher degree of differential rotation results in an increase of the additional viscosity needed during the subgiant phase. This is not the case as shown by the blue line in Fig.~\ref{rap_omegacs_seb_B}. A rapid decrease of the core rotation rate is indeed observed at the beginning of the post-main sequence due to the inclusion of the additional viscosity. Consequently, the ratio of core to surface rotation rates of the model without additional viscosity on the main sequence (blue line) becomes rapidly similar to the values obtained by using the opposite extreme assumption of solid-body rotation on the main sequence (black line). Only a very slight increase of the radial differential rotation is seen at the evolutionary stage of star B, which is completely negligible in view of the error bars on the observed rotation rates. We conclude that the internal transport of angular momentum during the main sequence has no impact on the determination of the efficiency of this transport during the subgiant phase.

Finally, we study the possible impact of the surface magnetic braking on the efficiency of the additional transport process needed during the post-main sequence. For this purpose, we compare models that share the same assumption for the internal transport of angular momentum but are computed with (green lines in Fig.~\ref{rap_omegacs_seb_B}) and without surface magnetic braking (black line). We see that the assumptions made for a possible surface magnetic braking have no impact on the determination of the efficiency of the internal transport of angular momentum during the subgiant phase. This only changes the initial velocity needed on the ZAMS in order to reproduce the surface velocity of post-main sequence stars. 

As for red giants, we conclude that asteroseismic measurements of subgiants can be used to determine the efficiency of the internal transport of angular momentum after the main sequence, independently from all the uncertainties of such a transport during the main sequence. This conclusion is of course valid for the six subgiants studied in this paper, but also for subgiants that are more evolved than star B. Figure~\ref{rap_omegacs_seb_B} suggests that this should also be the case for the majority of subgiant stars, except for young subgiants that are located close to the end of the main sequence (values of $\log(g)$ close to 4). As shown in Fig.~\ref{rap_omegacs_seb_B}, the rotational properties of the latter stars are indeed sensitive to the degree of radial differential rotation on the main sequence. These young subgiants constitute therefore perfect targets to constrain the internal transport of angular momentum on the main sequence from asteroseismic observations of evolved stars.

\section{Change of the additional viscosity with the evolution during the subgiant phase}
\label{evonuadd}

The key result of a decreasing efficiency of the internal transport of angular momentum with the evolution during the subgiant phase has important consequences regarding the physical nature of the missing transport process. Indeed, this directly explains why a transport process whose efficiency only depends on the degree of radial differential rotation (or more precisely to the ratio of core to surface rotation rates) cannot account for the core and surface rotation rates observed for subgiants. As shown by \cite{spa16}, such a process is able to correctly reproduce the variation of the core rotation rate along the red giant branch, but fails to do so during the subgiant phase. Indeed, a viscosity that depends on the ratio of core to surface rotation rates is only able to increase during the subgiant phase due to the simultaneous contraction of the central layers and increase of the stellar radius \citep[see Figs. 3 and 4 of][]{spa16}. This is exactly the opposite trend as the one deduced in the present work from the asteroseismic determination of core and surface rotation rates of subgiant stars.

\begin{figure}[htb!]
\resizebox{\hsize}{!}{\includegraphics{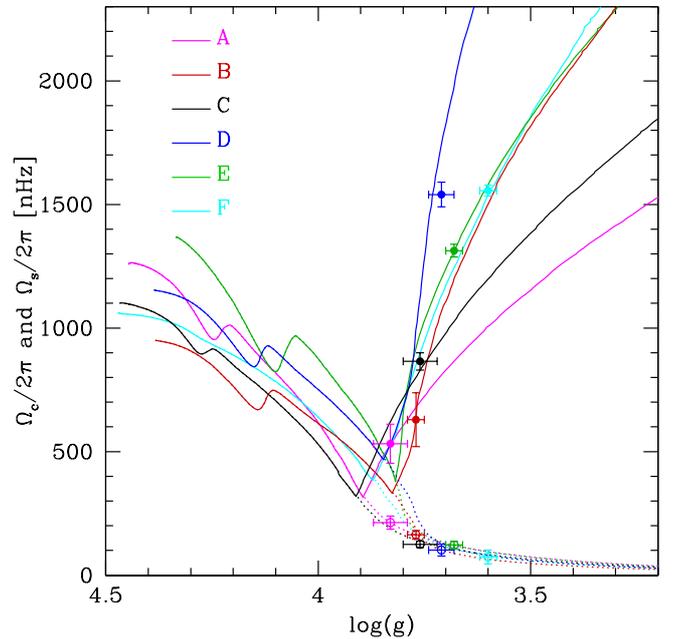}}
 \caption{Core and surface (solid and dotted lines) rotation rates as a function of gravity for rotating models of the six subgiants. These models are computed by assuming solid-body rotation until a given point during the post-main sequence. This decoupling point is directly determined by the need to correctly reproduce the observed core rotation rates. Dots and open circles indicate the values of core and surface rotation rates determined for the six subgiants from asteroseismic measurements \citep{deh14}.}
  \label{omgacs_vs_logg_seb_dec}
\end{figure}

While a decrease of the mean efficiency of the additional transport process is found during the subgiant phase, no clear constraints can be put on the way this decrease occurs. For instance, the efficiency of this transport mechanism can smoothly decrease during the whole subgiant phase, or can be very high just after the main sequence and abruptly tends to zero at a given point during the subgiant phase. The latter scenario was studied by \cite{spa16} by assuming the extreme case of solid-body rotation during the beginning of the post-main sequence evolution. Rigid rotation is then enforced until an arbitrary point during the subgiant phase, which is determined in order to best reproduce the observed core rotation rates. Using a fixed mass of 1.25\,$M_{\odot}$ and a solar chemical composition, \cite{spa16} find that enforcing solid-body rotation during the first 1\,Gyr of the post-main sequence evolution leads to a global satisfactory agreement with the core rotation rates observed for the six subgiants. 

\begin{figure}[htb!]
\resizebox{\hsize}{!}{\includegraphics{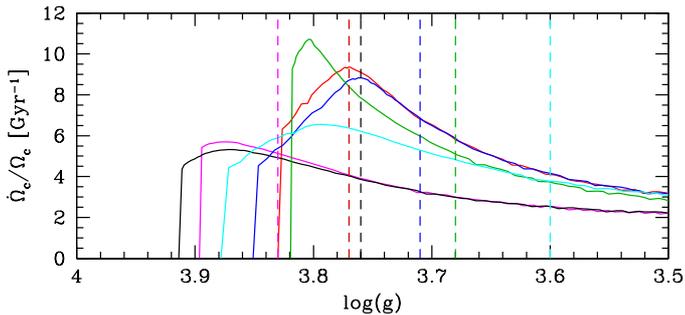}}
 \caption{Variation rate of the core angular velocity $\Omega_{\rm c}$ as a function of gravity for rotating models of the six subgiants computed by assuming solid-body rotation until a given point during the post-main sequence. Magenta, red, black, blue, green and cyan lines correspond to stars A, B, C, D, E and F, respectively. Vertical dashed lines indicate the values of the gravity for the six stars.}
  \label{domegac_omc_dt_seb_dec}
\end{figure}

We study this scenario by taking into account the differences in masses and chemical composition of the six subgiants. Models are then computed for each of the six subgiants by assuming solid-body rotation until a given point during the post-main sequence evolution, and then the transport of angular momentum in radiative zones is assumed to be solely due to meridional circulation and shear instability (without any additional transport process). The basic idea is then to investigate whether such a decoupling event occurs at a similar evolutionary stage for the six subgiants and try to identify a possible physical reason responsible for it. The core and surface rotation rates for these rotating models of the six subgiants are shown in Fig.~\ref{omgacs_vs_logg_seb_dec}. The corresponding evolution of the variation rate of the core angular velocity is plotted in Fig.~\ref{domegac_omc_dt_seb_dec} for these models. As the core rotation rate decreases when solid-body rotation is assumed (except during the overall contraction phase at the end of the main sequence, see Fig.~\ref{omgacs_vs_logg_seb_dec}), the variation rate of the core angular velocity is negative (and thus not seen in Fig.~\ref{domegac_omc_dt_seb_dec}) until the decoupling point is reached. After this point, the solid-body hypothesis is relaxed and the variation rates of the core angular velocity recover therefore the values obtained for rotating models computed without additional viscosity, which are shown in Fig.~\ref{domegac_omc_dt_seb}.

When differences in masses and chemical composition of the six subgiants are neglected, Fig.~\ref{omgacs_vs_logg_seb_dec} shows that the increase of the core rotation rate when the surface gravity decreases is globally reproduced by a rotating model without additional transport process after the decoupling. This is typically the case for the model corresponding to star B, which is characterized by a mass of 1.27\,$M_{\odot}$ and an approximately solar metallicity (red lines in Fig.~\ref{omgacs_vs_logg_seb_dec}). This is in good agreement with the results of \cite{spa16}, which show that core rotation rates of subgiants are correctly reproduced when solid-body rotation is enforced during the beginning of the post-main sequence phase. As shown in Fig.~\ref{omgacs_vs_logg_seb_dec}, an efficient additional mechanism must however rapidly operate at the end of the subgiant phase to prevent the large increase of the core rotation rates that would be in contradiction with the slow rotation of the central layers observed during the red giant phase.

When an individual modelling is performed for each of the six subgiants, Fig.~\ref{domegac_omc_dt_seb_dec} shows that the decoupling always takes place before the peak in the variation rate of the core angular velocity during the subgiant phase (when the decoupling occurs $\dot{\Omega}_{\rm c}/\Omega_{\rm c}$ passes from a negative to a positive value in Fig.~\ref{domegac_omc_dt_seb_dec}). However, this decoupling can occur just before this peak (see for instance star E, green line) or well before this peak is reached (as for star D, blue line). For the six subgiants, the decoupling cannot therefore be clearly associated with the same characteristic moment in the rotational evolution of the star. It thus appears that there is no clear physical reason to justify the hypothesis of a sudden decoupling during the subgiant phase. However, the general idea of an efficient transport mechanism during the main sequence whose efficiency gradually decreases after it, is favored by the present finding of a decreasing additional viscosity during the subgiant phase. This conclusion is in good agreement with the results of \cite{pin17} who show that internal gravity waves generated by penetrative convection could be a valuable candidate to reproduce such a behavior during the subgiant phase. 

As the efficiency of the internal transport of angular momentum must then rapidly increase with the evolution during the red giant phase, this may suggest that the physical nature of the additional transport process is different in main sequence and red giant stars. The trend during the red giant phase could be for instance reproduced by a transport mechanism whose efficiency depends on the degree of radial differential rotation \citep{spa16}. These preliminary findings need of course to be confronted to asteroseismic observations of evolved stars sharing different masses and evolutionary stages, with the ultimate goal of revealing the physical nature of the missing transport mechanism(s).

\section{Conclusion}
\label{conclusion}

Based on the asteroseismic determination of core and surface rotation rates obtained for six subgiant stars by \cite{deh14}, we investigated the efficiency of the transport of angular momentum in radiative zones during the subgiant phase. This was done by computing rotating models for each of the six subgiants and by determining the additional viscosity (in addition to the transport of angular momentum by meridional currents and shear instability) needed to correctly account for these asteroseismic constraints.

We first find that the efficiency of the internal transport of angular momentum can be precisely determined for each of the six subgiants thanks to the precise asteroseismic measurements of both core and surface rotation rates. The additional viscosities obtained for the six subgiants lie between $(4\pm2) \times 10^{3}$\,cm$^2$\,s$^{-1}$ and $(1.7\pm0.6) \times 10^{4}$\,cm$^2$\,s$^{-1}$. Compared to previous determinations, the efficiency of the transport of angular momentum is thus found to be higher in these subgiants than in the low-mass red giant KIC~7341231, but lower than in the more massive red giant KIC~8366239. 

We show that the efficiency of the transport of angular momentum during the subgiant phase can be determined independently from the rotational history of the star. In particular, this post-main sequence transport efficiency is insensitive to the uncertainties related to the modelling of the transport of angular momentum and surface magnetic braking during the main sequence. By recalling that the same result is obtained for red giants, we conclude that asteroseismic measurements of evolved stars provide us with a direct access to the evolution of the internal transport of angular momentum during the post-main sequence. The drawback of this conclusion is of course that we cannot used these data to constrain the transport of angular momentum in main sequence stars. Young subgiants (typical values of $\log(g)$ close to 4) constitute however an interesting exception in this context. Contrary to more evolved stars, the rotational properties of these young subgiants are indeed found to be sensitive to the internal transport of angular momentum on the main sequence. Asteroseismic observations of these stars would thus be particularly valuable to shed a new light on the transport of angular momentum in the radiative zones of main sequence stars.    

We also find that the efficiency of the undetermined additional process for the internal transport of angular momentum increases with the mass during the subgiant phase. The same trend is shown for more evolved red giant stars. The increase of the transport efficiency with the stellar mass seems therefore to be seen for evolved stars at different evolutionary stages; it remains to be studied whether this is also the case for stars in the core helium burning phase.    

We finally studied the change of the efficiency of the transport of angular momentum when evolution proceeds during the subgiant phase. We find that the efficiency of the internal transport of angular momentum decreases with the evolution of the star during the subgiant phase. This result obtained for subgiants is in clear contrast with the increase in the efficiency of the transport mechanism needed when the star ascends the red giant branch. This leads to interesting consequences regarding the physical nature of the missing transport process. In particular, the decrease in the efficiency of the transport of angular momentum during the subgiant phase implies that a transport mechanism whose efficiency increases with the degree of radial differential rotation cannot account for the core rotation rates of subgiant stars. We recall that such a process predicts an increase in the transport efficiency that correctly reproduces the core rotation rates of stars on the red giant branch \citep{spa16}. This may suggest that the physical nature of the undetermined transport process is different in subgiant and red giant stars. In this context, the decrease in the transport efficiency with the evolution found during the subgiant phase could be attributed to the gradual disappearance of an efficient transport process that was already at work during the main sequence. Then another efficient mechanism must be rapidly at work after the subgiant phase to correctly reproduce the core rotation rates of red giants. This possible scenario is of course very speculative, since we are still far from having a clear understanding of the internal transport of angular momentum. We will therefore continue our efforts to characterize in detail the additional transport mechanism needed after the main sequence thanks to asteroseismic observations of evolved stars, and thereby try to reveal its physical nature.

\begin{acknowledgements}
This work has been supported by the Swiss National Science Foundation (project Interacting Stars, number 200020-172505). A.M. and G.B. acknowledge support from the ERC Consolidator Grant funding scheme (project ASTEROCHRONOMETRY, G.A. n. 772293). N.L. acknowledges financial support from "Programme National de Physique Stellaire" (PNPS) and the "Programme National de cosmologie et Galaxie" (PNCG) of CNRS/INSU, France.
\end{acknowledgements}

%-------------------------------------------------------------------

\bibliographystyle{aa} % style aa.bst
\bibliography{biblio} % your references Yourfile.bib

\end{document}